\documentclass{ws-ijmpa}

\begin{document}

\markboth{F. Cianfrani, G. Montani}
{Curvature-spin coupling from the semi-classical limit of the Dirac equation}

%
\catchline{}{}{}{}{}
%

\title{Curvature-spin coupling from the semi-classical limit of the Dirac equation
}

\author{FRANCESCO CIANFRANI}

\address{ICRA---International Center for Relativistic Astrophysics\\ 
Dipartimento di Fisica (G9),
Universit\`a  di Roma, ``Sapienza'',\\
Piazzale Aldo Moro 5, 00185 Rome, Italy.\\ 
francesco.cianfrani@icra.it}

\author{GIOVANNI MONTANI}

\address{ICRA---International Center for Relativistic Astrophysics\\ 
Dipartimento di Fisica (G9),
Universit\`a  di Roma, ``Sapienza'',\\ 
Piazzale Aldo Moro 5, 00185 Rome, Italy.\\ 
ENEA C.R. Frascati (Dipartimento F.P.N.),
Via Enrico Fermi 45, 00044 Frascati, Rome, Italy.\\
ICRANet C. C. Pescara, Piazzale della Repubblica, 10, 65100 Pescara, Italy.\\
montani@icra.it}

\maketitle

\begin{history}
\received{Day Month Year}
\revised{Day Month Year}
\end{history}

\begin{abstract}

The notion of a classical particle is inferred from Dirac quantum fields on a curved space-time, by an eikonal approximation and a localization hypothesis for amplitudes. This procedure allows to define a semi-classical version of the spin-tensor from internal quantum degrees of freedom, which has a Papapetrou-like coupling with the curvature. 


\end{abstract}

\ccode{PACS number: 03.65.Sq, 04.62.+v}

\section{Introduction}

The definition of quantum fields on a curved background is an hard task, because of the impossibility to define a unique (up to unitary equivalence) Hilbert space structure \cite{BD84}. Nevertheless, equations of motion for such fields can be extended to the curved case and semi-classical tools can be applied at least formally. This way, the interaction of the gravitation field with quantum degrees of freedom is inferred, in a regime where Quantum Gravity effects are not yet relevant and a particle-like description is recovered. This coupling may be tested experimentally by studying the trajectories of elementary particles.

Here we address the point whether the interaction between quantum spin and the geometry reflects that of classical spin tensors. In fact, from the semi-classical limit of spinor fields, we expect a non-trivial coupling between their internal degrees of freedom and the curvature, so that a non-geodesics dynamics is expected. These deviations with respect to the structure-less case are going to be compared with those ones coming from the Papapetrou formulation for spinning particles \cite{pap}. 

The work is organized as follows: at first the Papapetrou formulation for classical spinning particles is presented, and then the semi-classical limit of the Dirac equation is performed.

\section{Papapetrou formulation}

Papapetrou \cite{pap} performed the classical analysis on the motion of a body in General Relativity on a fixed background.

In his formulation, the body is described in the space-time as a thin tube inside which there is a non-vanishing energy momentum tensor. As soon as a curve $X^\mu=X^\mu(\tau)$ is identified as the center-of-mass world-line, the multi-pole expansion consists in neglecting all moments of the energy-momentum distribution with respect to $X^\mu$ from a certain one. 
 
At the lowest order a geodesics dynamics is found and this is consistent with basic principles of General Relativity. At the next order, the pole-dipole approximation, a characterization of the body internal structure is given by the spin tensor $S^{\mu\nu}$, whose expression is the following one
\begin{equation}
S^{\mu\nu}=\int_{\tau}\delta x^{\mu}T^{\nu 0}-\int_{\tau}\delta x^{\nu}T^{\mu 0},
\end{equation} 

$\delta x^\mu=x^\mu-X^\mu$ being the deviation with respect to $X^\mu$. The system of equations describing the dynamics is
\begin{equation}
\left\{\begin{array}{c}\frac{D}{Ds}P^{\mu}=\frac{1}{2}R_{\rho\sigma\nu}^{\phantom1\phantom2\phantom3 \mu} S^{\rho\sigma}u^{\nu}\quad\\\\
\frac{D}{Ds}S^{\mu\nu}=P^{\mu}u^{\nu}-P^{\nu}u^{\mu}\\\\
P^{\mu}=mu^{\mu}-\frac{DS^{\mu\nu}}{Ds}u_{\nu}\quad\end{array}\right.\label{pap}
\end{equation}

and it outlines that for rotating bodies
\begin{itemize}
{\item a generalized momentum $P_\mu$ arises and, in general, it is not aligned with the 4-velocity;}
{\item the spin tensor is subjected to a precession around the generalized momentum;}
{\item the trajectory deviates with respect to a geodesics motion, as a consequence of the interaction between the spin tensor and the gravitational field.} 
\end{itemize}
 
However, in the system (\ref{pap}) there are thirteen unknown quantities and ten equations only. Hence, to provide a solution three conditions must be added. In literature one finds the following three possible choices, 
 \begin{equation}
S^{\mu 0}=0\qquad S^{\mu\nu}u_\nu=0\qquad S^{\mu\nu}P_\nu=0
\end{equation}
the Corinaldesi-Papapetrou \cite{pap2}, the Pirani \cite{pir} and the Tulczyjew \cite{tul} consistency conditions, respectively. All of them expresses the spatial nature of the spin.

\section{Semi-classical limit of the Dirac equation on a curved space-time}

The investigation on the coupling between quantum degrees of freedom and gravity requires the extension of the quantum fields formulation to a curved space-time.

As far as spin-$1/2$ particles are concerned, the Dirac algebra must be extended to non-Minkowskian spaces. This can be done by introducing vier-bein vectors $e^a_\mu$ and defining Dirac matrix as $\gamma_\mu=e^a_\mu\gamma_a$, $\gamma_a$ being those ones of the flat case. Furthermore, generalized derivatives must be defined \cite{BW57}, such that the Dirac equation in curved space-time reads as
 \begin{equation} 
i\hbar\gamma^\mu D_\mu\psi-m\psi=0,\qquad D_\mu\psi=\bigg(\partial_\mu-\frac{i}{2}\omega^{ab}_\mu\Sigma_{ab}\bigg)\psi,
\end{equation}
$\Sigma_{ab}=\frac{i}{4}[\gamma_a,\gamma_b]$ being generators of the Lorentz group, while $\omega^{ab}_\mu=e_\nu^b\nabla_\mu e^{a\nu}$.  

We are going to perform the semi-classical limit by the eikonal approximation, {\it i.e.}
\begin{equation}
 \psi=e^{iS}u
\end{equation}
$S$ being of the inverse Compton scale $(1/\lambda)$ order, while $u$ is a spinor.  

The notion of a classical trajectory is recovered for $u$ by a localization along the world-line $ x^\mu=x^\mu(\tau)$, which is the integral curve of $K_\mu=\partial_\mu S$. The localization is obtained by assuming a Gaussian dependence of the following kind

\begin{equation}
u(s,x)=\Pi_\mu\frac{e^{-\frac{(x^i-x^i(s))^2}{\sigma_i}}}{\sqrt{2\pi}\sigma_i}u_0
,\qquad \sigma_i\approx\lambda.
\end{equation}
 
Even though the spread $\sigma_i$ grows during the notion, nevertheless it can be shown that for a reasonable set of initial conditions there exists a macroscopic region where a classical trajectory can be defined \cite{io}.
 
The dynamics is inferred from the squared Dirac equation, {\it i.e.}
\begin{equation}
D^\mu D_\mu\psi-\frac{1}{4}R\psi+m^2\psi=0.
\end{equation}

In order to make this formulation closer to the Papapetrou one, we are going to perform a spacial integration, taking as space-like coordinates $x^i$, {\it i.e.} those ones of the space orthogonal to $K_\mu$.

The localization here works by allowing us to neglect any term containing first and second derivatives of the Gaussian. Furthermore, in agreement with the assumption that any dependence on spatial coordinates has been included into the Gaussian, we take for the spinor $u_0$
\begin{equation}
D_\mu u_0=i u_\mu v,
\end{equation}

$v$ being a spinor.
  
In this approximation scheme one obtains   
 
\begin{equation}
\int\bigg[\frac{1}{2}(\bar{\psi}\gamma^\nu D^\mu D_\mu\psi+h.c.-\frac{1}{4}R\bar{\psi}\gamma^\nu\psi+\mu^2\bar{\psi}\gamma^\nu\psi\bigg]e_\nu^0 d^3x\sqrt{-g}=(K_\mu K^\mu-K^\mu S_\mu)(1+O(\lambda^2))+m^2=0,\label{dr}
\end{equation}

which is the require dispersion relation.

$S^\mu$ is the first correction in the semi-classical limit and its expression is $S_\mu=-\frac{1}{2}\omega^{ab}_\mu e_{a\rho}e_{b\sigma}S^{\rho\sigma}$, $S^{\mu\nu}$ being the spin density, {\it i.e.}

\begin{equation}
S^{\mu\nu}=\frac{\int d^3x\sqrt{h}\bar{u}\{\gamma^{\bar{0}},\Sigma^{\mu\nu}\}u}{2\int d^3x\sqrt{h}\bar{u}\gamma^{\bar{0}} u}=\frac{\bar{u}_0\{\gamma^{\bar{0}},\Sigma^{\mu\nu}\}u_0}{2\bar{u}_0\gamma^{\bar{0}}u_0}+O(\lambda^2).
\end{equation}

It is worth noting that $S^{\mu\nu}$ satisfies Pirani consistency conditions $u_\nu S^{\mu\nu}=0$.
 
Hence, because of the relation

 \begin{equation}
\nabla_\nu S_\mu=\nabla_\mu S_\nu-\frac{1}{2}R_{\rho\sigma\mu}^{\phantom1\phantom2\phantom3\nu}S^{\rho\sigma}-2i\frac{D_{[\nu}\bar{u}_0\gamma^{\bar{0}}D_{\mu]} u_0}{\bar{u}_0\gamma^{\bar{0}}u_0},
\end{equation}

by acting with a covariant derivative $\nabla_\nu$ on the dispersion relation (\ref{dr}), the dynamics turns out to be described by the following system of equations  

\begin{eqnarray*}
\left\{\begin{array}{c}u^\mu\nabla_\mu P_\nu-\frac{1}{2}R_{\rho\sigma\mu\nu}u^\mu S^{\rho\sigma}=0\\
\frac{D}{Ds}S^{\mu\nu}=0\\
P_\nu=K_\nu-S_\nu=K_\nu-\hbar u_\nu(\bar{v}\gamma^{(0)}u+\bar{u}\gamma^{(0)}v)\end{array}\right..
\end{eqnarray*}

We stress that a Papapetrou-like interaction between the spin tensor and the Riemann tensor comes out. Furthermore, being the generalized momentum aligned with the 4-velocity, the conservation of the spin tensor is consistent with a Papapetrou-like dynamics.  

Therefore, the notion of particle coming out from the semi-classical limit of a spinor field coincides with a classical spinning particle, with spin tensor of the $\hbar$ order.


\end{document}